\documentstyle[sprocl]{article}

\def\be{\begin{equation}}
\def\ee{\end{equation}}
\def\bea{\begin{eqnarray}}
\def\eea{\end{eqnarray}}

\def\[{\left [}
\def\]{\right ]}
\def\({\left (}
\def\){\right )}

\begin{document}

\title{Observational Tests of Instanton Cosmology}

\author{R.E. Allen, J. Dumoit, and A. Mondragon}

\address{Center for Theoretical Physics, Texas A\&M University\\ 
College Station, Texas, 77843 USA \\ e-mail: allen@tamu.edu}


\maketitle\abstracts{A new cosmological model leads to testable predictions that are
different from those of both standard cosmology and models with a
cosmological constant. The prediction that $q_{0}=0$ is the same as in other
``coasting universe'' models, but arises without the need for any
exotic form of matter or other {\it ad hoc} assumptions.}

A new cosmological model was recently proposed, in which the big-bang
singularity and global metric of spacetime are due to a cosmological
instanton.~\cite{allen1,allen2,turok} The starting point is the order parameter
$\Psi _{s}$ for a GUT Higgs field which has a symmetry group 
SO(10) $\times$ SU(2) $\times$ U(1).~\cite{allen1} Since the coupling
to other fields is not discussed in the present paper, the SO(10) gauge 
symmetry will be ignored. $\Psi _{s}$ can then be treated simply as an 
SU(2)$\times$ U(1) order parameter, and there are two natural point defects:

(1) SU(2) instantons, which are the four-dimensional generalization
(in Euclidean spacetime) of two-dimensional vortices in 
ordinary superfluids. With the order parameter written in the form 
$\Psi _{s}=n_{s}^{1/2}U\eta _{0}$ and the scalings\\ 
$\rho =r/\xi$, $f=\left( n_{s}/\overline{n}_{s}\right)^{1/2}$, 
$\xi =\left( 2m\mu \right)^{-1/2}$, $\overline{n}_{s}=\mu /b$
in a standard notation,~\cite{fetter} the condensate density 
$n_{s}$ satisfies the generalized Bernoulli equation~\cite{allen1,allen2} 
\begin{equation}
-\frac{1}{\rho ^{3}}\frac{d}{d\rho }\left( \rho ^{3}\frac{df}{d\rho }\right)
+\frac{3a^{2}}{\rho ^{2}}f-f+f^{3}=0 \label{1}
\end{equation}
in Euclidean spacetime. The asymptotic solutions are 
\begin{equation}
f=1-3a^{2}/2\rho ^{2}\quad ,\quad \rho \rightarrow \infty \quad \quad \mbox{and}
\quad \quad  f=C\rho ^{n}\quad ,\quad \rho \rightarrow 0 \label{2}
\end{equation}
where $C$ is a constant and $n=\left( 1+3a^{2}\right) ^{1/2}-1.$ An $n=1$
BPST-like instanton then has a strength $a=1$.

(2) U(1) monopole-like defects, which act as sources of condensate current, 
and which have a Euclidean action~\cite{allen1,allen2}
\begin{equation}
S=\overline{n}_{s}\mu \int d^{4}x\,f^{\dagger}\left[ -\frac{1}{\rho ^{3}}\frac{d}{d\rho }
\left( \rho ^{3}\frac{df}{d\rho }\right)
-\frac{\overline{a}^{6}}{\rho ^{6}(f^{\dagger}f)^{2}}f-f+\frac{1}{2}(f^{\dagger}f)f\right].
\label{3}
\end{equation}
In the Euclidean formulation, $f$ is not necessarily real and 
$f^{\dagger}$ is not even necessarily 
the Hermitian conjugate of $f$. (See the comments below (2.10) and 
(3.12) of Ref. 1.) Here, however, we will be concerned 
only with the behavior of $\Psi_{s}$ in the Lorentzian spacetime of 
human observers, where $f$ is real and $n_{s}$ is positive.

A general point defect can have both $a$ and 
$\overline{a}$ nonzero. The cosmological instanton of Refs. 1 and 2 is such a
defect, and the radial coordinate in four-dimensional spacetime is
interpreted as the time coordinate: $x^{0}=r$. After a Wick rotation to 
Lorentzian spacetime, the density of the bosonic field 
$\Psi _{s}$ is determined by the equation 

\begin{equation}
\frac{1}{\rho ^{3}}\frac{d}{d\rho }\left( \rho ^{3}\frac{df}{d\rho }\right)
+\frac{3a^{2}}{\rho ^{2}}f-\frac{\overline{a}^{6}}{\rho ^{6}f^{4}}f
-f+f^{3}=0 \label{5}
\end{equation}
and the asymptotic solutions are 
\begin{equation}
f=1-3a^{2}/2\rho ^{2}\quad ,\quad \rho \rightarrow \infty \quad \quad \mbox{and}
\quad \quad  f=c\rho ^{-1}\quad ,\quad \rho \rightarrow 0. \label{6}
\end{equation}
Although $f$ diverges as $\rho \rightarrow 0$, there is a
cutoff at the Planck length $\ell _{P}$, \cite{allen1,allen2} which is
analogous to the cutoff at the atomic scale in an ordinary 
superfluid.

Let $\rho _{c}$ be defined by $f\left( \rho _{c}\right) =1$ and let 
$\overline{\eta}=\overline{a}/a$. Realistic values of $\overline{a}$ and 
$a$ are, of course, cosmologically large. For example, if $\overline{t}$
is the value of the time $t$ (in the Robertson-Walker metric) that corresponds
to $\rho =\overline{a}$, and if $\overline{t}$ is $\sim 100$ seconds, the
results of Ref. 2 imply that $\overline{a}\sim 10^{44}$.
Nevertheless, it is informative to solve the differential equation
(\ref{5}) numerically for more modest values, and Fig. 1 shows the solution for 
$\overline{a}=\overline{\eta }=10$. One expects that $f\left( \rho
\right)$ is slowly decreasing for $\overline{a}<\rho<\rho _{c}$, that 
$f\left( \rho \right) \approx 1$ in this same range, and that $\rho _{c}=
\overline{a}\left( \overline{a}^{2}/3a^{2}\right) ^{1/4}$ to a very good
approximation. These expectations are confirmed by the 
results of Fig. 1, and by calculations that we have performed for 
other values including $\overline{a}=\overline{\eta }=100$.

\begin{figure}[h]
\input{fig1.tex} 
\centerline {Fig. 1. Scaled value of $n_{s}^{1/2}$ versus
scaled value of $x^{0}$, for $a=1$ and $\overline{a}=10$.}
\end{figure}

In the present model, the universe is born at $\rho \sim 1$ (or $x^{0}\sim
t_{P}\sim 10^{-42}\sec $) with a highly confined and coherent bosonic field 
$\Psi _{s}$. Through interactions with other fields that are omitted in the
present paper, it releases a small portion of its initially very large
energy, producing a hot big bang. At the same time, it relaxes into a fully
condensed state during the period $0<\rho<\rho_{c}$. 
It then remains fully condensed, so that $n_{s}=\overline{n}_{s}$ for 
$\rho \ge \rho _{c}$. This requires that~\cite{allen2}
\begin{equation}
v_{\alpha }^{0}v_{\alpha }^{0}=v_{\alpha }^{k}v_{\alpha }^{k}\quad ,\quad
\rho >\rho _{c}\ \label{7}
\end{equation}
where
\begin{equation}
v^{\mu }=v^{\mu }_{\alpha}\sigma^{\alpha}
=-im^{-1}U^{-1}\partial ^{\mu }U \label{8}
\end{equation}
is the generalization of $m\overrightarrow{v}=\overrightarrow{\nabla }\theta $
in an ordinary superfluid. 

The condition (\ref{7}) has the form $\omega^{2}=p^{2}$,
where $p^{k}=mv^{k}$ describes the rotations of the
two-component order parameter $\Psi _{s}$ as a function of position. 
This condition can be written in a third
form: $v_{\alpha}^{\mu }v_{\mu }^{\alpha }=\eta _{\mu \nu }v_{\alpha }^{\mu} 
v_{\alpha }^{\nu }=0$,
where $\eta _{\mu \nu }=diag(-1,1,1,1)$ is the Minkowski metric tensor. 
When it is satisfied, the generalized Bernoulli equation
\begin{equation}
-\frac{1}{2m}n_{s}^{-1/2}\partial ^{\mu }\partial _{\mu }n_{s}^{1/2}+\frac{1%
}{2}mv_{\alpha }^{\mu }v_{\mu }^{\alpha }+bn_{s}=\mu \label{9}
\end{equation}
holds with $n_{s}=\overline{n}_{s}=\mu /b$. I.e., the GUT Higgs 
field $\Psi_{s}$ is fully condensed for large times $\rho>\rho_{c}$.

In the theory of Ref. 1,
the geometry of spacetime is determined by the field $v_{\alpha }^{\mu }$,
which is interpreted as the vierbein: $g^{\mu \nu }=\eta ^{\alpha \beta}
\,e^{\mu }_{\alpha }e^{\nu }_{\beta }$ with $e_{\alpha }^{\mu }=v_{\alpha
}^{\mu }$. For example, massless fermions and vector bosons move just as in
standard general relativity, except for a breaking of Lorentz invariance at
high energies, and a small additional gravitational interaction in the case
of fermions. The local curvature of spacetime results from Planck-scale
instantons, just as the rotation of a superfluid results from quantized
vortices. The Einstein field equations hold as a very good approximation for
local deformations of spacetime geometry, but the global metric of the
universe is determined by the  cosmological SU(2)$\times $U(1) instanton
described above. One obtains a metric with the usual Robertson-Walker 
form\\
$ds^{2}=-dt^{2}+R\left( t\right) ^{2}\left[ dr^{2}/(1-r^{2})+r^{2}
d\Omega \right]$
but with a cosmic scale factor which is quite different from that of
standard Friedmann models:~\cite{allen2}
\begin{equation}
R\left( t\right) =\eta \,t\quad ,\quad t\rightarrow 0 \label{10}
\end{equation}
\begin{equation}
R\left( t\right) \propto \left[ \int dt/n_{s}\left( t\right) \right]
^{1/2}\quad ,\quad t<t_{c} \label{11}
\end{equation}
\begin{equation}
dR\left( t\right) /dt=2\sqrt{3}\quad ,\quad t>t_{c} \label{12}
\end{equation}
where $\eta \approx 2\overline{\eta }\gg 1$. Recall that $n_{s}\left( 
t\right)$ is a decreasing function of $t$ and
$n_{s}(t)\approx \overline{n}_{s}$ for $\overline{t}<t<t_{c}$.
These forms follow from the solutions to (\ref{5}) and (\ref{7}), with the details
given in Ref. 2. Here $\overline{t}$ and $t_{c}$ are the times corresponding
to $\rho =\overline{a}$ and $\rho =\rho _{c}$ respectively. For example,  
$\overline{t}\sim 100$ sec and $\overline{\eta }\sim 100$ corresponds to 
$t_{c}\sim 10^{6}$ sec.

The behavior of the scale factor in (\ref{10})-(\ref{12}) has major 
implications:~\cite{allen2} (\ref{10}) and (\ref{12}) offer solutions to the 
smoothness, monopole, and flatness problems, and (\ref{11}) is fully 
consistent with big-bang nucleosynthesis. In addition, (\ref{12}) 
implies a comfortably large age for the universe, and predicts that 
the deceleration parameter $q_{0}$ is exactly zero. This is quite 
different from the predictions $q_{0}\geq 1/2$, for Friedmann models 
with a flat or closed universe, or $q_{0}\sim -1/2$, for
$\Lambda$CDM models.~\cite{turner1} There have been other proposals 
for a ``coasting universe'', but they involve exotic 
forms of matter~\cite{kolb1} or other {\it ad hoc} 
assumptions,~\cite{ozer}
whereas here the result $q_{0}=0$ 
emerges automatically within the context of a more general theory.

\section*{Acknowledgements}

This work was supported by the Robert A. Welch Foundation.

\end{document}